\documentstyle[aaspp4,epsfig]{article}
\newcommand{\bc}[1]{{\bf#1}}
\begin{document}
\title{Statistics of Weak Gravitational Lensing in Cold Dark
Matter Models; Magnification Bias on Quasar Luminosity Functions}

\author{Takashi Hamana,\altaffilmark{1} 
        Hugo Martel,\altaffilmark{2}
        and Toshifumi Futamase\altaffilmark{1}}

\altaffiltext{1}{Astronomical Institute, Tohoku University, 
                 Sendai 980-8578, Japan}
\altaffiltext{2}{Department of Astronomy, University of Texas, 
                 Austin, TX 78712}

\begin{abstract}
We compute statistical properties of weak gravitational
lensing by large-scale structure
in three Cold Dark Matter (CDM) models: two flat models with
$(\Omega_0,\lambda_0)=(1,0)$ and $(0.3,0.7)$ and one open model with
$(\Omega_0,\lambda_0)=(0.3,0)$, where $\Omega_0$ and $\lambda_0$ are
the density parameter and cosmological constant, respectively.
We use a Particle-Particle/Particle-Mesh (P$^3$M)
$N$-body code to simulate the formation and evolution
of large-scale structure in
the universe. We perform $1.1\times10^7$ ray-tracing experiments for each 
model, by computing the Jacobian matrix along random lines of sight,
using the multiple lens-plane algorithm.
From the results of these experiments, we calculate the probability
distribution functions of the convergences, shears, and magnifications, 
and their root-mean-square (rms) values.
We find that the rms values of the convergence and shear agree
with the predictions of a nonlinear analytical model.
We also find that the probability distribution functions of the 
magnifications $\mu$ have a peak at values slightly smaller than 
$\mu=1$, and are strongly skewed toward large magnifications. 
In particular, for the high-density ($\Omega_0=1$) model,
a power-law tail appears in the distribution function at large
magnifications for sources at redshifts $z_s>2$.
The rms values of the 
magnifications
essentially agree with the nonlinear analytical predictions for
sources at low redshift, but exceed these predictions for high redshift 
sources, once the power-law tail appears.

We study the effect of magnification bias on the luminosity functions of
high-redshift quasars, using the
calculated probability distribution functions of the magnifications. 
We show that the magnification bias is moderate in the absence of
the power-law tail in the magnification distribution, but depends
strongly on the value of the density parameter $\Omega_0$.
In presence of the power-law tail, the bias becomes
considerable, especially at the bright end of the luminosity functions
where its logarithmic slope steepens.
We present a specific example which demonstrates that the bias flattens
the bright side logarithmic slope of a double power-law luminosity
function.
\end{abstract}



\section{INTRODUCTION}

It is well known that the apparent brightness of distant sources are
gravitationally affected by the inhomogeneous distribution of 
matter in the universe, an effect called the lensing magnification.
Consequently, the apparent brightness and intrinsic
luminosity of a distant source are no longer simply related by the
luminosity distance-redshift relation in a smooth Friedmann universe
(Weinberg 1972; Schneider, Ehlers, \& Falco 1992, hereafter SEF).
The relation depends on how light rays are being lensed by 
density inhomogeneities in the universe, as they propagate from the 
source to the observer.
Since the pioneering works of Kristian \& Sachs (1966) and Gunn
(1967), numerous studies of the effect of lensing magnification in an
inhomogeneous universe have been published.
Two different approaches have been used in these studies to compute
the statistical properties of lensing magnifications in 
inhomogeneous universes with realistic matter distributions:
the power spectrum approach, which is analytical, and the numerical approach,
which combines $N$-body simulations of structure formation with
ray-tracing simulations.

The power spectrum approach was first developed by Gunn (1967), who
showed that the root-mean-square (rms) fluctuation in the apparent
brightness of distant sources (or equivalently in the lensing
magnification) can be expressed as a radial integral of the 
density autocorrelation function.
Gunn's original approach was eventually modified by Babul \& Lee (1991) 
to reflect substantial advances in our understanding of large-scale 
structure formation. These authors 
gave a quantitative estimate of the rms fluctuations
for modern cosmological models such as the Cold Dark Matter (CDM) model.
Frieman (1997) rederived the rms fluctuations using the Limber's equation in
Fourier space (Kaiser 1992, 1998), 
and found that the dispersion in the lensing magnification
can be larger than 0.1 for sources at redshift $z_s=1$, but depends
strongly on the background cosmological model.
The skewness of the probability distribution of the lensing magnifications 
was first calculated by Nakamura (1997) using the quasi-linear
theory of density fluctuation.

The first study based on the numerical approach
was done by Jaroszy{\'n}ski et al. (1990),
who used a Particle-Mesh (PM) $N$-body code to simulate the 
formation and evolution of large-scale structure in a CDM universe.
The multiple lens-plane algorithm developed by Blandford \& Narayan (1986) and
Kovner (1987) was then used to follow the evolution of ray bundles
propagating through the inhomogeneous matter distribution in the
simulated universes.
The force resolution of the PM simulations was $1h^{-1}\rm Mpc$,
in present units
(where $h$ is the Hubble constant $H_0$ in units
of $\rm100\,km\,s^{-1}Mpc^{-1}$), not sufficient to include lensing
effects by small-scale nonlinear structures like galaxy clusters.
They found that the rms values of the lensing magnifications by
large-scale structure are very small even for sources at redshift $z_s=5$.
Since then, this method has been improved by several authors
(Bartelmann \& Schneider 1991, 1992; Jaroszy\'{n}ski 1991, 1992;
Wambsganss et al. 1995, 1997; Wambsganss, Cen \& Ostriker 1998;
Premadi, Martel \& Matzner 1998; Tomita 1998a, 1998b).
Thanks to their effort and recent rapid developments of computational
techniques as well as computing power, the length resolution
of these algorithms has significantly improved.
For instance, Wambsganss et al. (1998) used large PM $N$-body
simulations combined with a convolution method and have 
achieved an effective resolution of $10 h^{-1}$kpc.

Jaroszy\'nski (1991, 1992) extended the length resolution of the method
to subgalactic scales by combining numerical simulations of large-scale
structure formation with a Monte-Carlo method for locating galaxies inside the 
computational volume. Galaxies were modeled as isothermal spheres
with velocity dispersions
and core radii determined by empirical relations depending on
morphological types. Ray-tracing simulations were then performed,
taking into account the combined effects of the large-scale structure
and the galaxies. Later, Premadi et al. (1998) improved this method
significantly, first by replacing the Zel'dovich algorithm used by
Jaroszy\'nski for simulating large-scale structure formation by
a more accurate Particle-Particle/Particle-Mesh (P$^3$M) $N$-body code, 
and then by taking into account, when assigning morphological types
to galaxies, the observed morphology-density relation
(Martel, Premadi, \& Matzner 1998 and references therein).

In this paper, we use the multiple lens-plane algorithm
to study the effect of lensing
by large-scale structure on quasar luminosity functions. 
We focus on CDM models with a {\sl COBE}-normalized
tilted power spectrum (Bunn \& White 1997). 
The P$^3$M simulations for these models
were provided by the Texas P$^3$M Database (Martel \& Matzner 1999).
Each model is characterized by the value of 4 parameters:
the density parameter $\Omega_0$, cosmological constant $\lambda_0$,
Hubble constant $H_0$, and rms density fluctuation $\sigma_8$
at scale $8h^{-1}\rm Mpc$ (the tilt $n$ of the power spectrum is a
dependent parameter). We consider three particular models,
an Einstein de Sitter model (E-dS)
with $\Omega_0=1$ and $\lambda_0=0$, an open model (O) with
$\Omega_0=0.3$ and $\lambda_0=0$, and a flat model
($\Lambda$) with $\Omega_0=0.3$ and $\lambda_0=0.7$.
For each model, we follow the propagation of light
ray bundles through the matter distribution. 
We are primarily interested in the statistical
properties of lensing by the matter inhomogeneities
in the universe, and not in rare events such as multiple imaging of
distant quasars. 
Consequently, we focus on the lensing effects by the large-scale
($>0.1h^{-1}$Mpc) structures and make no attempt to study lensing
effects caused by individual galaxies.

In order to obtain good statistics,
we perform a total of $1.1\times10^7$ ray-tracing experiments for each model.
Each experiment consists of
computing the Jacobian matrix along a random line of sight.
Having such a large number of experiments allows us to study in
detail not only rms values, but also the probability
distribution functions of the lensing properties, i.e., the
convergence, shear, and magnification.
We compare these rms values
with predictions based on the power spectrum approach.
This enables us to test the validity of the power spectrum approach
and its limitation.

Using the probability distribution functions of the lensing
magnifications obtained from the experiments, we study the effects of 
the magnification bias on quasar luminosity functions.
This problem was first studied by Sanitt (1971), and later by many
different authors (e.g., Turner 1980; Avni 1981; Canizares 1982; Peacock
1982; Vietri 1985; Ostriker \& Vietri 1986; Schneider 1987b, 1987c,
1992; see also SEF, chapter 12).
A crucial difference between these previous studies and ours is that we
compute the probability distribution in universes 
with realistic matter distributions, that originate from the growth of
primordial density fluctuations with a specific power spectrum
and normalization, and are simulated using a state-of-the-art $N$-body code.
In previous studies, the probability distribution was
calculated by considering simple matter
distributions such as randomly distributed point masses and/or
isothermal spheres.
These studies revealed that the magnification bias 
can have a considerable effect on the luminosity functions, 
especially for bright quasars (Ostriker \& Vietri 1986; Schneider 1992).
However, they cannot be used to make predictions about the
effect on the magnification bias in specific cosmological models, since
this requires a realistic representation of the large-scale structure
in the universe.

We demonstrate the importance of the bias on quasar luminosity functions and
show its dependence on the mean matter density of the universe.
However, we do not attempt to derive a precise quantitative 
estimate of the bias on the observed luminosity functions.
This would require a precise knowledge of both the observed luminosity
functions and the probability distribution of the lensing magnifications. 
However,
uncertainties in the quasar luminosity functions are still
considerable (e.g., Boyle, Shanks, \& Peterson 1988; Hartwick \& Schade 1990;
Warren, Hewett, \& Osmer 1994; Hawkins \& V\'eron 1995; 
La Franca \& Cristiani 1997).
Furthermore, the magnification distributions we calculate 
are those for a point like source, the effect of finite source size
on the magnification probability becomes important at large 
magnifications  (Schneider 1987b; Schneider \& Weiss 1988a, 1988b).
Because of these limitations, we decided to approximate the
intrinsic quasar luminosity
functions using either a single or a double power-law model.
Although it is not clear whether the power-law models provide an 
accurate description of a real luminosity function, we believe that
these models are sufficiently realistic (especially the double power-law
model) for a qualitative study of the essential properties of the
magnification bias.

The remainder of this paper is organized as follows: 
In \S2, we describe the cosmological models, and the
numerical methods used for simulating both
the large-scale structure formation and the light propagation. 
In \S3, we present the results of the experiments and compared them with
predictions based on the power spectrum approach.
In \S4, we discuss the effects of magnification bias on the quasar 
luminosity functions. Summary and conclusion are presented in \S5. 
In Appendix A, we summarize the basic equations used 
in the power spectrum approach.

\section{DESCRIPTIONS OF THE COSMOLOGICAL MODELS AND NUMERICAL METHODS}

\subsection{Cold Dark Matter Models}

\begin{deluxetable}{ccccccc}
\tablecaption{Summary of model parameters \label{table1}}
\tablewidth{0pt}
\tablehead{
\colhead{Model} & \colhead{$\Omega_0$} & \colhead{$\lambda_0$} & 
\colhead{$\Omega_{B0}h^2$} & \colhead{$h$} & \colhead{$\sigma_8$} & 
\colhead{$n$}
}
\startdata
E-dS      & 1.0 & 0.0 & 0.015 & 0.65 & 1.2 0& 0.8506 \nl
O         & 0.3 & 0.0 & 0.015 & 0.75 & 0.85 & 1.1748 \nl
$\Lambda$ & 0.3 & 0.7 & 0.015 & 0.75 & 0.90 & 0.8796 \nl
\enddata
\end{deluxetable}

We consider three tilted CDM cosmological models. Each model is
characterized by its values of $\Omega_0$, $\lambda_0$, $H_0$, and $\sigma_8$,
with the primordial exponent $n$ being a dependent parameter
(Martel \& Matzner 1999; Premadi et al. 1999).
Table~1 lists the models and gives their parameters.
The power spectrum is given by
\begin{equation}
\label{power_sp}
P(k)=2\pi^2\biggl({c\over H_0}\biggr)^{3+n}\delta_H^2k^nT^2(k)\,,
\end{equation}

\noindent where $c$ is the speed of light. The transfer function $T(k)$
for CDM models is given by Bardeen et
al.\ (1986) as follows,
\begin{equation}
\label{BBKS}
T(q)={\ln(1+2.34q)\over2.34q}[1+3.89q+(16.1q)^2+(5.46q)^3+(6.71q)^4]^{-1/4}\,,
\end{equation}

\noindent where $q$ is defined by
\begin{eqnarray}
\label{alpha_HS}
q&=&\biggl({k\over{\rm Mpc^{-1}}}\biggr)\alpha^{-1/2}(\Omega_0h^2)^{-1}
\Theta_{2.7}^2\,,\\
\alpha &=& a_1^{-\Omega_{\rm B0}/\Omega_0}
a_2^{-(\Omega_{\rm B0}/\Omega_0)^3}\,,\\ 
a_1 &=& (46.9 \Omega_0 h^2)^{0.670} \left[ 1 + (32.1 \Omega_0
h^2)^{-0.532} \right]\,,\\
a_2 &=& (12.0 \Omega_0 h^2)^{0.424} \left[ 1 + (45.0 \Omega_0
h^2)^{-0.582} \right]\,,
\end{eqnarray}

\noindent (Hu \& Sugiyama 1996), where $\Theta_{2.7}$ is the cosmic microwave
background temperature in units of 2.7K, and $\Omega_{\rm B0}$
is the contribution of baryons to the density parameter. In all
models, $\Theta_{2.7}$ and $\Omega_{\rm B0}$ were set equal
to 1 and $0.015h^{-2}$ respectively.
The density perturbation $\delta_H$ at horizon crossing is
obtained by fitting the {\sl COBE} 4-year data (Bunn \& White 1997),
as follows,
\begin{equation}
10^5\delta_H=\cases{
1.95\Omega_0^{-0.35-0.19\ln\Omega_0-0.17\tilde n}e^{-(\tilde n+0.14\tilde n^2)}
\,,& $\lambda_0=0$;\cr
1.94\Omega_0^{-0.785-0.05\ln\Omega_0}e^{-(0.95\tilde n+0.169\tilde n^2)}\,,
& $\lambda_0=1-\Omega_0$;\cr
}
\end{equation}

\noindent where $\tilde n\equiv n-1$.

\subsection{The P${}^3$M Algorithm}

The simulations of large-scale structure formation were provided by
the Texas P$^3$M Database (Martel \& Matzner 1999).
All simulations were performed using a P$^3$M $N$-body code
(Hockney \& Eastwood 1988) with $64^3$ 
particles, in a computational cubic box with triply periodic boundary conditions. The forces on particles are computed by solving
Poisson's equation on a $128^3$ cubic lattice using a Fast Fourier Transform
method. The forces at short distance are corrected by direct summation over
pairs of particles separated by less than some cutoff distance $r_e$
equal to a few grid spacings. The algorithm reproduces accurately
the Newtonian interaction between particle
down to the softening length $\eta$, which is set equal to a fraction of 
the grid spacing. In all simulations, the comoving lengths of
the size $L_{\rm box}$ of the computational box and
the softening length $\eta$ were equal to $128\,\rm Mpc$ and
$300\,\rm kpc$, respectively. Hence, the algorithm has a dynamical range of 
427 in length. Notice that in this paper, 
we express distances in units of $h^{-1}\,\rm Mpc$. In these units, the values
of $L_{\rm box}$ and $\eta$ vary among models.
These values, and also the mass per particle, are listed in Table~2.
Three simulations were performed for each cosmological model.
For each model, the simulations differ only in the
choice of random phases in the initial conditions.
All simulations start at an initial redshift of $z=24$, and end at
$z=0$.

\begin{deluxetable}{lccc}
\tablecaption{Parameters in P$^3M$ simulations \label{table2}}
\tablewidth{0pt}
\tablehead{
\colhead{$ \hbox{\strut Model}                   \atop $} & 
\colhead{$ \hbox{\strut Box Size, $L_{\rm box}$} \atop 
           \hbox{($h^{-1}$Mpc)}                        $} &
\colhead{$ \hbox{\strut Particle mass}           \atop 
           \hbox{($h^{-1}M_\odot$)}                    $} & 
\colhead{$ \hbox{\strut Softening length, $\eta$} \atop 
           \hbox{($h^{-1}$Mpc)}                        $}
}
\startdata
E-dS            & 83.2 & $3.608\times10^{11}$ & 0.195 \nl
O and $\Lambda$ & 96.0 & $1.249\times10^{11}$ & 0.225 \nl
\enddata
\end{deluxetable}

\begin{figure}
\begin{center}
\begin{minipage}{8.5cm}
\begin{center}
\epsfig{figure=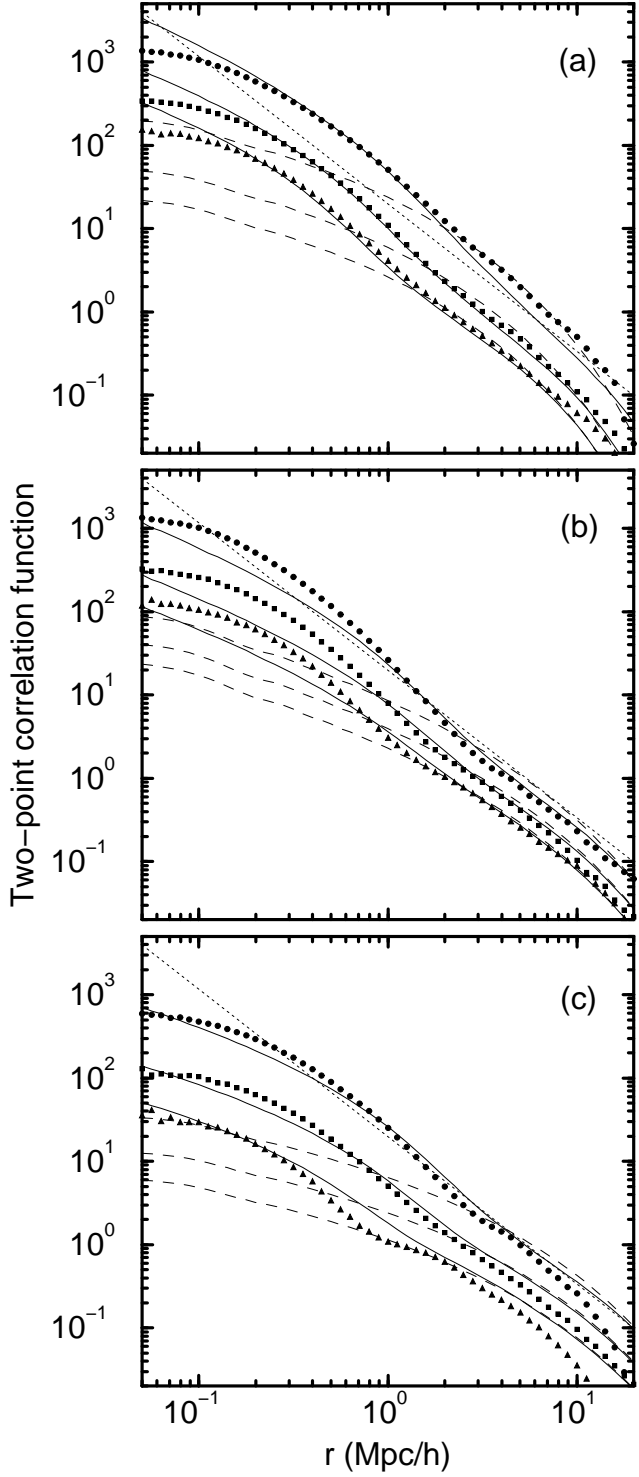,width=8.5cm}
\end{center}
\end{minipage}
\hspace{0.2cm}
\begin{minipage}{7.5cm}
\begin{center}
\caption[]{
Two-point correlation functions compared with those
predicted by the linear theory and 
(a) is for E-dS model, (b) is for O model and (c) is for
$\Lambda$ model.
The filled symbols represent the two-point correlation functions
calculated from the particle positions in the P${}^3$M simulations.
Filled circles, squares, and triangles corresponds to redshifts
$z=0$, $z=1$, and $z=2$, respectively.
Solid lines show the correlation function derived from the
non-linear power spectrum, using the fitting formula by Peacock \& Dodds 
(1996). Dashed lines show the correlation functions predicted by
linear theory. The redshifts are $z=0$,
$1$ and $2$ from top to bottom, respectively. The dotted line
shows the observed galaxy two-point correlation function.\label{fig1}}
\end{center}
\end{minipage}
\end{center}
\end{figure}

We calculated the two-point correlation function of the particles at redshifts
$z=0$, $z=1$, and $z=2$ using a direct estimator 
(Hockney \& Eastwood 1988),
\begin{equation}
\label{ch6-2cor}
\xi (r) = {{N_p} \over {\bar{n} N_c  \delta V}} -1,
\end{equation}

\noindent
where $N_p$ is the number of particles inside a spherical shell of
inner radius $r-\Delta r/2$ and outer radius $r+\Delta r/2$ centered
around a central particle, $\delta V$ is the volume of
this shell, $N_c$ is the number of particles taken as centers, and
$\bar{n}$ is the mean number density of the particles. 
The results are plotted in Figure \ref{fig1}. 
For comparison,
we plot analytical estimates of the two-point correlation
function, based both on linear perturbation theory and on
the nonlinear power spectrum fitting formula of Peacock \& Dodds (1996), 
and also the observed galaxy 2-point correlation function at the
present epoch,
\begin{equation}
\xi(r)=\biggl({r\over5.4\,h^{-1}{\rm Mpc}}\biggr)^{-1.77}
\end{equation}

\noindent (Peebles 1993).
As we see in Figure \ref{fig1} the slope the correlation
functions drop significantly at short range, for $r<\eta$.
Martel (1991) pointed out that this flattening
of $\xi$ is due to the softening of the force, which reduces the number of
particle pairs with separations $r\lesssim\eta$, and increases the
number of pairs with separations $r\gtrsim\eta$.
For the E-dS model, the correlation functions obtained from the P${}^3$M
simulation are in good agreement with the nonlinear predictions.
For the O and $\Lambda$ models, the nonlinear predictions are slightly
smaller than the correlation functions obtained from the simulations
at intermediate pair separations ($\eta < r <1h^{-1}$Mpc).
In \S2.3, we examine the statistical properties of 
the gravitational lensing by these simulated matter distributions, and 
compare the rms values of the
lensing convergences, shears, and magnifications with the
predictions of the power spectrum approach using Peacock \& Dodds'
fitting formula for the nonlinear power spectrum.
Therefore, the reader should keep in mind that, for the O and $\Lambda$
models, Peacock \& Dodds' fitting formula slightly underestimates the
correlation function at short range, $r<1h^{-1}$Mpc, and thus we can
expect that the rms values of the convergences, shears, and magnifications
predicted by the power spectrum approach will be also underestimated. 

\subsection{The Ray-Tracing Experiments}

We use the multiple-lens plane algorithm to follow
the evolution of light rays traveling through the simulated matter
distributions.
The P$^3$M simulations provided particle distributions in cubic boxes
at various redshifts, from the initial redshift $z=24$ to to present.
By combining these boxes, we can represent the matter distribution
in the universe along a line of sight extending from the observer to
the source.\footnote{This approach ignores the possible existence
of structures larger than $L_{\rm box}=128\rm\,Mpc$.
This does not effect on our experiments in any significant way,
because in the CDM universe, fluctuations at that scale have
a very small amplitude even at the present time, 
and their lensing effect is negligibly small.} 
For sources at redshift $z_s=3$, this requires 
36 boxes for the E-dS model, 38 for the O model, and 47 for
the $\Lambda$ model.
In order to eliminate spurious correlations between the large-scale
structure in adjacent boxes, we combine boxes from different simulations, 
so that cubic boxes which are directly joined to each other are chosen from
different simulations.

In the standard multiple-lens plane algorithm, the matter content of each
box is projected onto a single plane perpendicular to the line of sight.
We increase the accuracy of the algorithm by 
dividing the cubic box into four rectangular subboxes of
size $L_{\rm box} \times L_{\rm box} \times (L_{\rm box}/4)$, and
projecting the matter content of each subbox onto a plane, thus increasing
the number of lens planes by a factor of 4.

The deflection potential $\psi^i$ on the $i$-th lens plane is related to the
surface mass density fluctuation $\delta\Sigma_i(\vec{x})=\Sigma_i(\vec{x}) 
-\langle \Sigma_i \rangle$ on that plane by
\begin{equation}
\label{ch6-barpsi}
\psi^i (\vec{x}) = 
{1\over\pi}\int\!\!\!\int d^2 x'\,\delta\Sigma_i(\vec{x}')
\ln{|\vec{x}- \vec{x}'| \over x_0},
\end{equation}

\noindent
where $\vec x=(x_1,x_2)$ is the position vector in the plane, and
$x_0$ is an arbitrary cutoff length (its value is irrelevant, since only
derivatives of $\phi^i$ have a physical signification). This equation can be
rewritten in the form of a two-dimensional Poisson equation,
\begin{equation}
\label{fish}
\nabla^2 \psi^i = 2 \delta \Sigma_i\,.
\end{equation}

We solve this equation numerically on each lens plane by first computing the
surface density on a $512\times512$ square lattice from the particle
positions, using the Triangular Shaped Cloud (TSC) assignment scheme
(Hockney \& Eastwood 1988, \S5.3), and then inverting 
equation (\ref{fish}) using a Fast Fourier
Transform method (see, e.g., Premadi et al. 1998).
Notice that the grid spacings are $0.1625h^{-1}$Mpc for the E-dS
model and $0.1875h^{-1}$Mpc for the O and $\Lambda$ models, and are
slightly smaller than the softening length $\eta$ of the P$^3$M calculations.
The evolution equation of the Jacobian matrix in the
multiple-lens plane algorithm is given by
\begin{equation}
\label{ch6-mle_barpsi}
\bc{A}_{j+1} = \bc{I}-{{4 \pi G} \over {c^2}} 
\sum_{i=1}^j {D_i D_{ij} \over D_j} \bc{U}_i \bc{A}_i,
\end{equation}

\noindent
where $D_j$ is the angular diameter distance
between the observer and the $j$-th lens plane,
$D_{ij}$ is the angular diameter distance
between the $i$-th, $j$-th lens plane,
and $\bc{U}_i$ is an optical tidal matrix defined by,
\begin{equation}
\label{ch6-barU}
\bc{U}_i
=\left(
\begin{array}{cc}
{\psi}_{,11}^i&
{\psi}_{,12}^i\\
{\psi}_{,12}^i &
{\psi}_{,22}^i
\end{array}
\right)
=\left(
\begin{array}{cc}
\delta\Sigma_i + {1 \over 2}({\psi}_{,11}^i-
{\psi}_{,22}^i)&
{\psi}_{,12}^i\\
{\psi}_{,12}^i &
\delta\Sigma_i - {1 \over 2} ({\psi}_{,11}^i-
{\psi}_{,22}^i)
\end{array}
\right)
\,,
\end{equation}

\noindent
with commas denoting differentiation with respect to $\vec{x}$.
The last equality was obtained using equation (\ref{fish}).
We compute the second derivatives of $\psi^i$ on grid points
using a standard finite difference formula.

In general, equation (\ref{ch6-mle_barpsi}) is not an explicit equation for
${\bc{A}}_i$ since the summation contains the tidal matrix
$\bc{U}_i$, which must be evaluated along the light ray path. 
One first has to solve the multiple gravitational lens equation in
order to compute the location of the light ray on each lens plane
and then evaluate $\bc{U}_i$ at that location. 
However, the deflections of light rays are very small; the
deflection angle caused by structures larger than $0.1h^{-1}$Mpc
is at most $\alpha\sim10^{-5}$.
The distance between the actual and unperturbed positions of a light ray
will be much smaller than the grid spacing of the
lattice used for calculating the
two-dimensional potential and its derivatives.
To give an actual example, 
the transverse distance from the unperturbed ray position is at most
$l_{\perp} \sim L_{\rm box} \alpha < 10^{-3}h^{-1}$Mpc for a light ray
propagating a comoving distance of $L_{\rm box}$.
We can therefore treat light rays as if they were moving along straight 
lines.\footnote{We could not make this simplification 
if we were interested in the
deformation and/or multiple imaging of sources with a finite size. In that
case, it is the relative difference between the deflection angles of
nearby rays that matters, and those can be quite large.}
This greatly simplifies the algorithm. Not only
the computation of light ray trajectories becomes unnecessary, 
but in addition
we can choose lines of sights that go through the grid points, 
where the surface mass density fluctuation $\delta \Sigma_i$ and the
derivatives $\psi^i_{,11}$, $\psi^i_{,22}$ and
$\psi^i_{,12}$ of the potential are known,
thus eliminating the need to interpolate $\bc{U}_i$
from the grid points to the location of the rays. Since the
lattice is composed of $512\times512=262\,144$ grid points, there are
$262\,144$ possible lines of sight to choose from. Each experiments consists
of selecting one line of sight, and solving equation (\ref{ch6-mle_barpsi})
recursively along that line of sight for all values of $\bc{A}_j$,
starting with $\bc{A}_1=\bc{I}$ (the identity matrix). 
It may seem that the number of possible experiments is limited to
$262\,144$, but actually, since the P$^3$M simulations use periodic boundary
conditions, each cubic box can be given a random shift, or,
equivalently, any grid point in a cubic box can be
chosen, independently of the ones chosen in other cubic boxes,
as long as the same grid point is chosen on the four lens planes
within each box.
The total number of possible experiments is therefore $262\,144^{N_{\rm box}}$,
where $N_{\rm box}$ is the number of cubic boxes along the line of sight.

We decompose the Jacobian matrix as 
\begin{equation}
\label{Ch6-Adecomp}
\bc{A} = 
\left(
\begin{array}{cc}
1-\delta \kappa -\gamma_1 & -\gamma_2 -\omega\\
-\gamma_2 +\omega & 1-\delta \kappa +\gamma_1 
\end{array}
\right),
\end{equation}

\noindent
where $\delta \kappa$ represents a fluctuation of convergence from
that in smooth Friedmann universe, $|\gamma|=(\gamma_1^2+\gamma_2^2)^{1/2}$ 
is the amplitude of shear of a light ray bundle, and $\omega$ is the
rotation angle of the beam.
The image magnification factor of a point like source is given by the
inverse of the determinant of the Jacobian matrix, 
\begin{equation}
\mu={1 \over |\det \bc{A}|}
={1 \over\left|(1-\delta \kappa)^2 -\gamma^2 -\omega^2 \right|}.
\end{equation}

\section{RESULTS}

\subsection{Lensing Convergence and Shear}

\begin{figure}
\begin{center}
\begin{minipage}[t]{8cm}
\begin{center}
\epsfig{figure=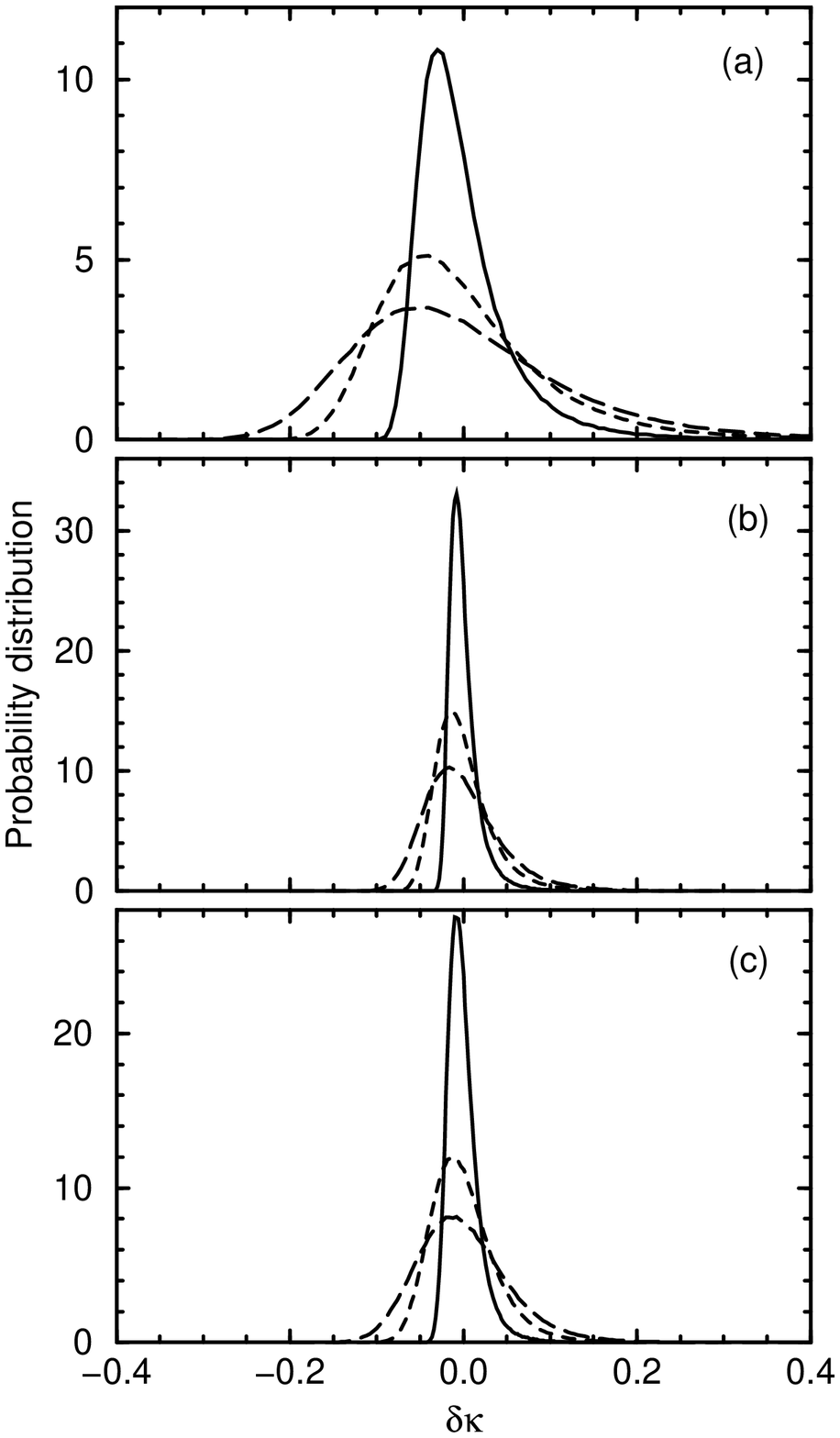,width=8cm}
\caption[]{
Probability distributions of fluctuations of lensing convergence,
$\delta \kappa$.
The solid lines, short-dashed lines, and 
long-dashed lines correspond to sources at redshifts $z_s=1$,
2, and 3, respectively.
(a) E-dS model; (b) O model; (c) $\Lambda$ model.
\label{fig2}}
\end{center}
\end{minipage}
\hspace{0.2cm}
\begin{minipage}[t]{8cm}
\begin{center}
\epsfig{figure=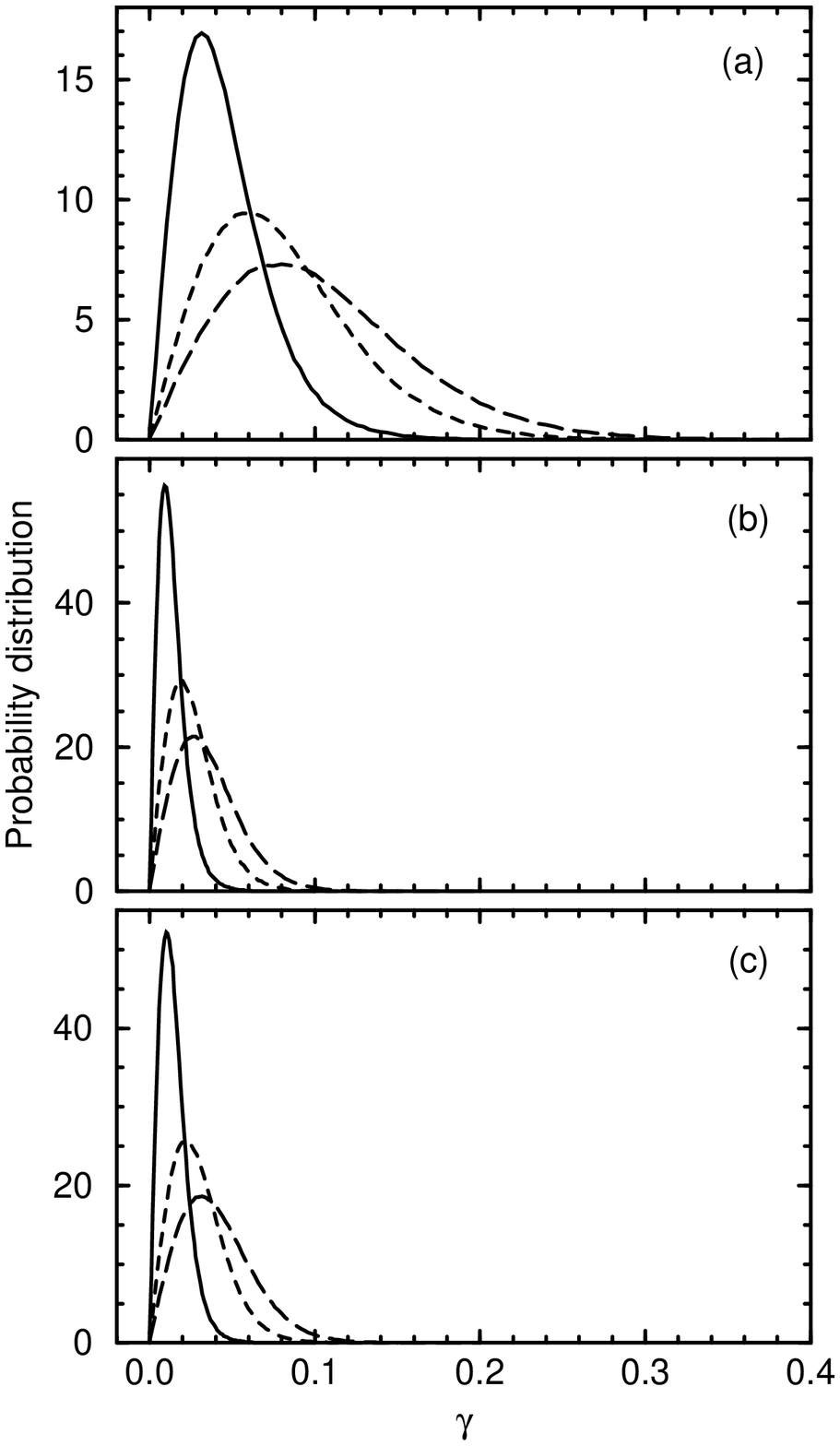,width=8cm}
\caption[]{
Probability distributions of the lensing shear of images.
Lines have the same meaning as in Figure \ref{fig2}.
(a) E-dS model; (b) O model; (c) $\Lambda$ model.
\label{fig3}}
\end{center}
\end{minipage}
\end{center}
\end{figure}

We first study the statistical properties of the lensing
convergence $\delta\kappa$ and shear $\gamma$.
For this purpose we perform $10^6$ experiments for each model. 
The probability distributions of $\delta \kappa$ and $\gamma$ are shown 
in Figure \ref{fig2} and \ref{fig3}, respectively, for
sources at redshifts $z_s=1$, 2, and 3.
The probability distributions of both $\delta\kappa$ and $\gamma$
broaden when the source redshift increases, and
those in the E-dS model (top panel) are considerably broad compared with
those in the O (middle panel) and $\Lambda$ (bottom panel) models.
The probability distributions are slightly broader
in the $\Lambda$ model than in the O model.
These properties are consistent with analytical predictions,
as we will demonstrate below. The probability distributions
of $\delta\kappa$ have a peak at some negative value,
and are skewed toward positive values.
This behavior is prominent in the E-dS model. 
These distributions are a reflection of the mass distribution in the 
universe: in most of the universe, the matter density is
smaller than the average value, while most of the matter is
concentrated into compact structures such as galaxies and clusters of
galaxies. 
Thus, for the majority of the random lines of sight, light ray bundles 
converge less compared to those propagating through a smooth Friedmann
universe. The average values of the fluctuations of the
lensing convergence in our simulations are found to be consistent with
zero, i.e., $|\langle \delta \kappa \rangle| < 10^{-4}$. 
There is a minimum value of the fluctuation of the lensing
convergence, which occurs when the beam propagates through empty space.
It is given by 
\begin{equation}
\label{del_kappa_min}
\delta \kappa_{\min} = 1 -{{D_{DR}(z)} \over {D_F(z)}},
\end{equation}

\noindent (SEF, chapter 11; Hamana 1998),
where $D_{DR}(z)$ is the so-called Dyer-Roeder distance (Dyer \&
Roeder 1972, 1973) 
with the clumpiness parameter $\tilde{\alpha}=0$ (Fukugita et al.\ 1992), 
and $D_F(z)$ is the standard angular diameter distance.
The values of $\delta\kappa_{\min}$ are given in Table~3 for
the three cosmological models considered in this paper.
Comparing the lower cutoff of the probability distributions found in
Figure \ref{fig2} with these minimum values, we find that a value 
of $\delta\kappa$ near $\delta \kappa_{\min}$
is unlikely to occur in our ray-tracing experiments,
especially for sources at a high redshift. This is consistent with the 
results of Futamase \& Sasaki (1989), which are 
based on a simple analytical approach.

\begin{deluxetable}{cccc}
\tablecaption{The minimum values of the fluctuation of the lensing
convergence, $\delta \kappa_{\min}$ \label{table3}}
\tablewidth{0pt}
\tablehead{
\colhead{Redshift of source $z_s$} & \colhead{E-dS} & 
\colhead{O} & \colhead{$\Lambda$}
}
\startdata
1.0      & $-0.1243$ & $-0.04504$ & $-0.06567$ \nl
2.0      & $-0.3285$ & $-0.1274$  & $-0.2071$  \nl
3.0      & $-0.5500$ & $-0.2203$  & $-0.3724$  \nl
\enddata
\end{deluxetable}

\begin{figure}[t]
\begin{center}
\begin{minipage}{16cm} 
\epsfig{figure=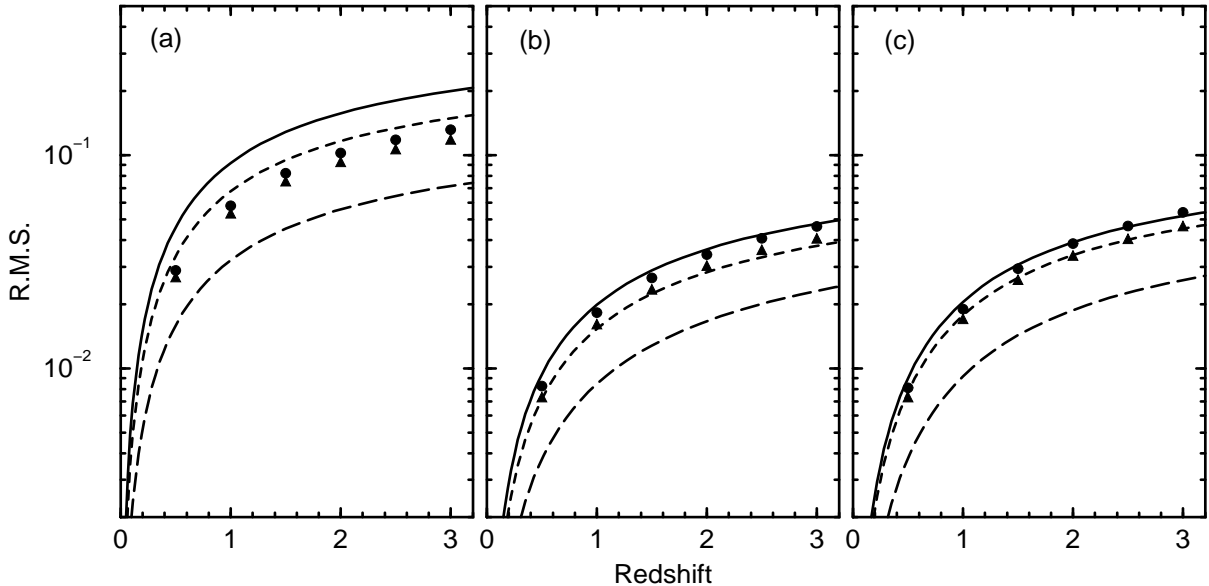,width=16cm,angle=-90}
\end{minipage}
\end{center}
\caption[]{
Root-mean-squares of the lensing convergence and shear of
images as a function of redshifts of source.
The filled circles and filled triangle are for the lensing convergence
and shear of images evaluated from the results of ray-tracing
experiments, respectively.
The solid curves represent predictions of the
power spectrum approach with the nonlinear power spectrum by Peacock
\& Dodds (1996), dashed curves are those obtained from a top-hatted
density field with a smoothing scale of $R_s=0.1h^{-1}$Mpc, and
long-dashed curves are the predictions of linear perturbation theory.
(a) E-dS model; (b) O model; (c) $\Lambda$ model.
\label{fig4}}
\end{figure}

In Figure \ref{fig4}, we plot the rms values $\sigma_\kappa$ of 
the convergence and $\sigma_\gamma$ of the shear, defined by
\begin{eqnarray}
\sigma_{\kappa} &=& \left[ {1 \over {N_{\rm exp}}} \sum_{i=1}^{N_{\rm exp}} 
(\delta\kappa_i)^2 \right]^{1/2},\\
\sigma_{\gamma} &=& \left[ {1 \over {N_{\rm exp}}} \sum_{i=1}^{N_{\rm exp}} 
| \gamma_i |^2 \right]^{1/2},
\end{eqnarray}

\noindent
where $N_{\rm exp}$ is the number of experiments.
Notice that $\sigma_\kappa$ and $\sigma_\gamma$ are
expected to be equal, based on analytical arguments (Jain \& Seljak 1997).
Instead, we find that 
the rms values of the shear computed from our ray-tracing experiments are
slightly but systematically smaller than those of the convergence.
However, this difference is probably 
insignificant, because the shear computed from the experiments 
might be underestimated, whereas the convergence is probably accurate.
The shear is mainly caused by the
traceless symmetric part  (Weyl part) of the optical tidal matrix
$\bc{U}_i$, which depends upon
second derivatives of the two-dimensional potential $\psi^i$.
These second derivatives can be underestimated somewhat because of
the finite resolution of the lattice used for solving equation (\ref{fish}).
The lensing convergence is
mainly caused by the surface mass density fluctuation $\delta\Sigma_i$,
which is directly computed from the particle positions.

In each panel of Figure \ref{fig4}, we also plotted three curves,
which represent various analytical estimates of $\sigma_\kappa$
and $\sigma_\gamma$ (see Appendix A); the solid curves
represent predictions based on the power spectrum approach 
(Kaiser 1992; Bernardeau, van Waerbeke \& Mellier 1997; Jain \& Seljak 1997;
Nakamura 1997), using the
nonlinear power spectrum by Peacock \& Dodds (1996) 
(hereafter, we refer to this as the nonlinear method), dashed curves
represent the predictions obtained from a top-hat filtered density
field with a smoothing scale of $R_s=0.1h^{-1}$Mpc (the filtered
nonlinear method), comparable to the softening length of the
P${}^3$M simulations, and long-dashed curves represent
the predictions of linear perturbation theory (the linear method).

The nonlinear method predicts that the rms values scale approximately as
$\sigma_{\kappa},\sigma_{\gamma} \propto \sigma_8 \Omega_0^{\sim0.7-0.8}$ 
at a fixed source redshift, and increase with the cosmological constant
because of the resulting increase in the path length
(Bernardeau et al. 1997; Jain \& Seljak 1997). 
In all models, the rms values obtained from our ray-tracing experiments
are significantly larger than the predictions of the linear method.
These rms values should be
compared preferentially to the predictions of the filtered nonlinear
method, because the effect of lensing by structures smaller scale than
the softening length is not included in our experiments.
The rms values obtained from our experiments are slightly smaller than
those predicted by the filtered nonlinear method for the E-dS model,
and slightly larger for the O and $\Lambda$ models.
It is important to recall that the same trends were shown in the
two-point correlation functions of the matter (Figure
\ref{fig1}).
These two trends are consistent with each others at least
qualitatively; the correlation function and the rms values are either
both overestimated or both underestimated by the nonlinear
predictions, depending upon the cosmological model.  
This suggests that the rms values obtained from our
experiments are consistent at least qualitatively
with the predictions of the filtered nonlinear method.

These various comparisons show that rms values of the lensing
convergence and shear are identical within the uncertainties inherent
to the numerical method. They also show 
that the power spectrum approach with the
nonlinear power spectrum provides good estimates for the
rms values of the lensing convergences and shears.

\subsection{Lensing Magnification}

Next, we study the statistics of the lensing magnifications.
In order to obtain meaningful statistics, and be able to study properties
of light rays with a large magnification, which are
very rare among random lines of sight, we perform $10^7$ 
experiments for each model. 

\begin{figure}[t]
\begin{center}
\begin{minipage}{16cm}
\epsfig{figure=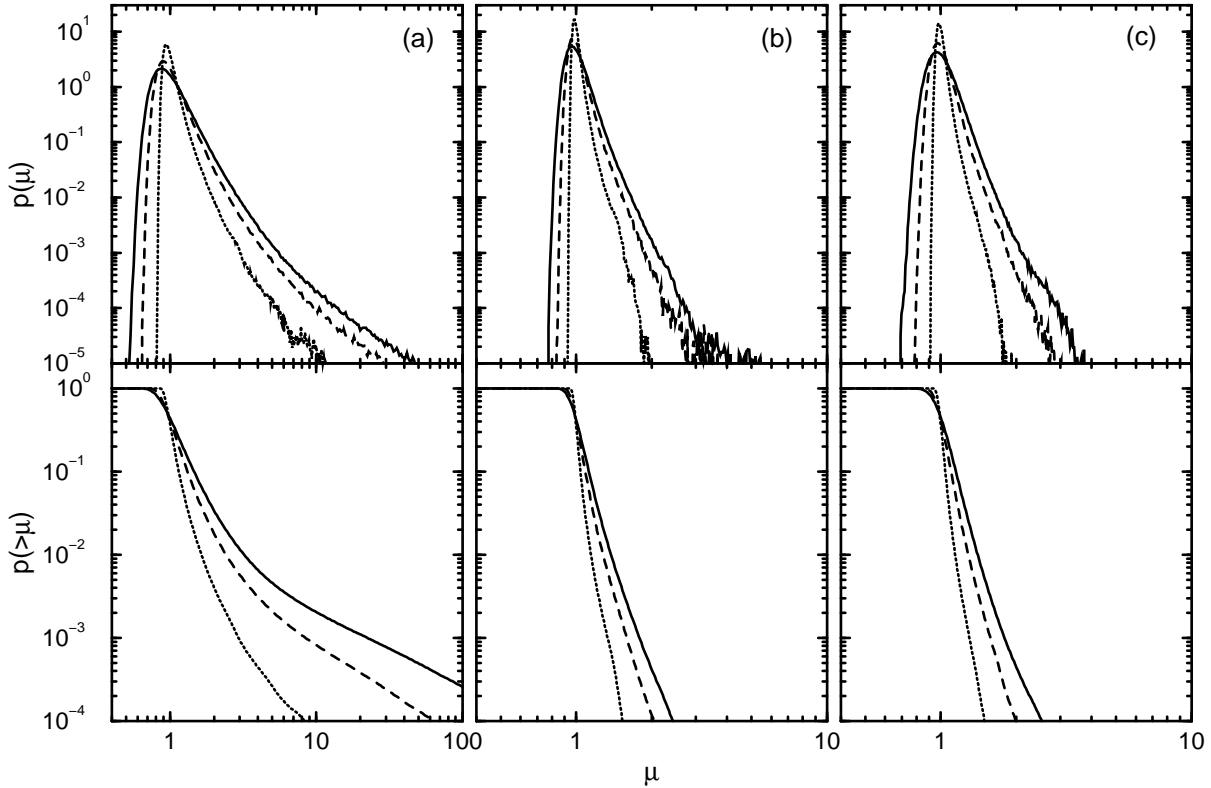,width=16cm,angle=-90}
\end{minipage}
\end{center}
\caption[]{
Probability distributions of image magnifications.
The solid lines are for the source redshift $z_s=3$, dashed lines are
for $z_s=2$, and dotted lines are for $z_s=1$.
(a) E-dS model; (b) O model; (c) $\Lambda$ model.
For each model, top panel shows the differential distribution
$p(\mu)$, and bottom panel shows the integral one $p(>\mu)$. 
\label{fig5}}
\end{figure}

In Figure \ref{fig5}, we plot the probability distributions of image
magnifications of a point-like source.
For each model, the top panel shows the differential distribution
$p(\mu)$, and the bottom panel shows the integral distribution $p(>\mu)$.
The probability distributions are normalized so as to satisfy the
normalization constraint $\int_0^{\infty} d \mu\,p(\mu)=1$.
In the differential distributions, we find that the distributions have
a strong peak located at values of $\mu$ slightly less than 1,
and are significantly skewed toward large magnifications. 
In the case of the E-dS model, (Figure \ref{fig5}a), a 
power-law tail of the form $p(\mu)\propto \mu^{-2}$
appears at high magnifications, for sources located at redshifts $z_s=2$ and 3.
This power-law behavior is a general property of the lens equation for 
a point-like source (see, e.g., SEF, chapter 11).
The appearance of this power-low tail indicates the existence of
caustics.

The magnification distributions shown in Figure \ref{fig5} can be
interpreted as the ``transfer function'' of the matter in the universe 
(Wambsganss et al.\ 1998).
In other words, any intrinsic luminosity function of, say, quasars
will be folded with this magnification distribution, and what we measure
as the observed quasar luminosity function will be the convolution of
the intrinsic luminosity function with this transfer function of the
universe (Turner 1980; Avni 1981; Canizares 1982, Peacock 1982; Vietri
\& Ostriker 1983; Vietri 1985; Ostriker \& Vietri 1986; Schneider 1992).
Therefore, it is essential to study first the magnification
distribution in order to compute the magnification bias accurately.
We shall discuss this point in detail in \S4. 

Since gravitational lensing not only
magnify sources but also causes a distortion of their area on the sky,
the lines of sight used in our experiments, which are randomly
distributed on the {\it image plane} are not
randomly distributed on the {\it source plane} (Ehlers \& Schneider 1986).
The small area of a source on the source plane is enlarged by a factor of
$\mu$ on the image plane.
Therefore if one averages the lensing magnifications over images
randomly distributed on the {\it image plane}, the resulting average
magnification $\langle\mu\rangle$ will be larger than unity.
Taking the effect of area distortion into account, the
probability distribution of the lensing magnifications of sources,
i.e., the probability distribution calculated from the magnifications
of sources randomly distributed on a source plane, is given by
\begin{equation}
\label{ps(mu)}
p_s(\mu) = {p(\mu)\over\mu}\,.
\end{equation}

\noindent
Notice that, by definition, the mean magnification over sources automatically
becomes unity,
\begin{equation}
\label{fluxconv}
\langle \mu \rangle = \int_0^{\infty} d\mu\,\mu p_s(\mu) 
= \int_0^{\infty} d\mu\,p(\mu) =1\,. 
\end{equation}

\begin{figure}[t]
\begin{center}
\begin{minipage}{16cm} 
\epsfig{figure=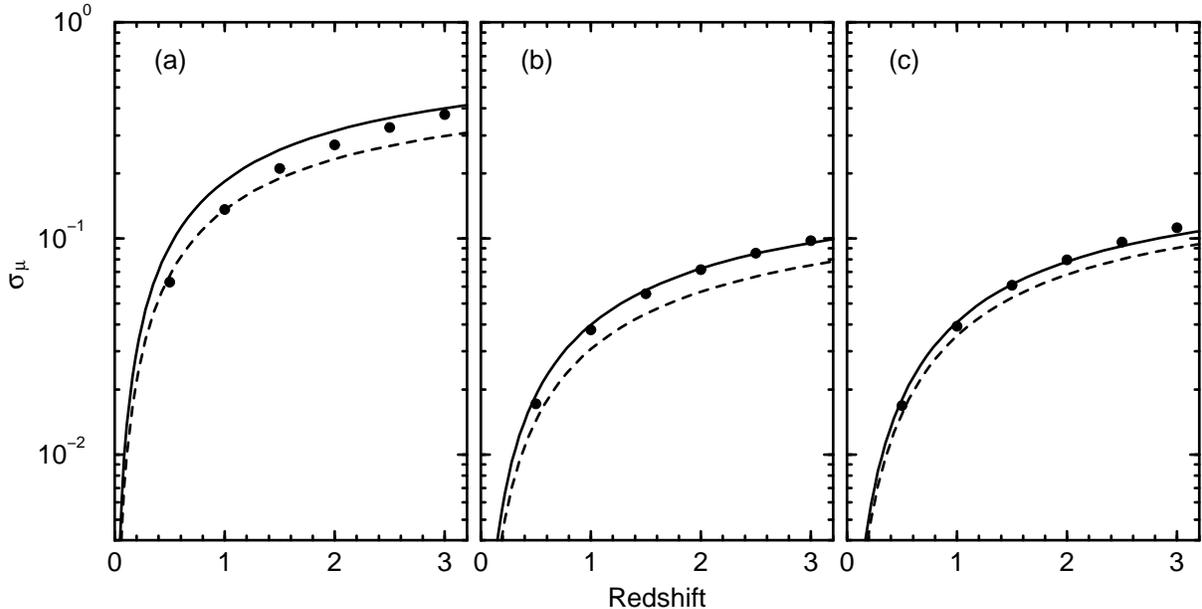,width=16cm,angle=-90}
\end{minipage}
\end{center}
\caption[]{
The dispersions in the lensing magnification of sources $\sigma_{\mu}$ 
versus source redshift.
The filled circles are evaluated from the results of the ray-tracing
experiments.
The solid curves represent predictions of the nonlinear method,
dashed curves are those of the filtered nonlinear method.
(a) E-dS model; (b) O model; (c) $\Lambda$ model.
\label{fig6}}
\end{figure}

\noindent
Using the probability distributions shown in
Figure \ref{fig5}, we calculated the dispersions of the magnification
distributions,
\begin{equation}
\label{mu_stand_disp}
\sigma_{\mu}^2 = \int_0^\infty d\mu \left( \mu -1 \right)^2 p_s(\mu)
\simeq \sum_i \left( \mu_i -1 \right)^2 p_s(\mu_i)\,,
\end{equation}

\noindent
where the probability distribution functions $p_s(\mu)$
are sampled in bins equally spaced in intervals of 0.01 in $\log \mu$, 
in the range $0.1\leq\mu\leq100$.
The results are shown as filled circles in Figure \ref{fig6}.
The solid curves represent the predictions of the power spectrum
approach, using the nonlinear power spectrum of Peacock \& Dodds (1996),
and the dashed curves represent predictions 
based on the top-hat filtered density
field with a smoothing scale of $R_s=0.1h^{-1}$Mpc.
The dispersions 
in the E-dS model are larger than those in the O and $\Lambda$ models by
about a factor of 3. In all models, the magnification
dispersion is larger than 0.1 for a source redshift $z_s=3$. 
The effect of magnification dispersion on
observations of sources will clearly be important 
for such high redshifts, especially if the density parameter 
$\Omega_0$ is large.
 
Figures 6b and 6c show that, for the O and $\Lambda$ models, the
dispersions obtained from the experiments are in good agreement
with the predictions
of the nonlinear method, but are larger that the predictions of the
{\it filtered} nonlinear method, as in the cases of lensing
convergences and shears.  
This is again consistent with the trend found in the two-point
correlation function at least qualitatively; i.e., the fitting formula 
of the power spectrum of Peacock \& Dodds (1996)
underestimates the two-point 
correlation functions for the O and $\Lambda$ models.
Figure \ref{fig4}a shows that for the E-dS model, 
the rms values agree with the predictions of the filtered nonlinear method 
when the source redshift is smaller than 1.
However, the growth of $\sigma_\mu$ with the source redshift
is more rapid than the filtered nonlinear method predicts.
This rapid growth reflects the appearance of the power-law tail in the
magnification distribution.
In the power spectrum approach, a perturbative treatment is adopted
for calculating the lensing magnification, and thus high
magnification events are not properly taken into account.
Therefore, the dispersion of
lensing magnification estimated from the power spectrum approach should
be regarded as a lower limit.

Notice that the integral in equation (\ref{mu_stand_disp})
diverges once the power-law behavior at the high magnification appears. 
In our calculations, this divergence is prevented by the
fact that we have a finite number of experiments: the probability $p_s$
eventually drops to zero at some value of $\mu$, simply because no
experiments have produced larger values.
In the real universe, the integral does not
diverge because the power-law tail is effectively cut off at large
magnifications. This is caused by the following two effects.
First, astrophysical sources are extended, and their magnifications
(given by the surface brightness-weighted point-source magnification
over the solid-angle area of the source) 
remain finite (e.g. Bontz 1979; Schneider 1987a; see also SEF, chapter 12).
Second, even a point source can only be magnified by a finite value, because
the geometrical optics approximation 
is inadequate near critical curves. A more accurate
treatment, based on a wave-optics description of lensing, shows that
the magnification is always finite 
(e.g., Ohanian 1974; 1983; Bliokh \& Minakov 1975; Nakamura 1998;
also see SEF, chapter 7).

We should note here that, as was mentioned above, the rms values are
sensitive to the normalization of the density contrast field,
$\sigma_8$, i.e.\ roughly $\sigma_{\mu} \propto \sigma_\kappa,\sigma_\gamma
\propto \sigma_8$ (Bernardeau et al. 
1997; Jain \& Seljak 1997; Nakamura 1997).
Thus the large rms values and the appearance of the power-law tail in
magnification distribution in E-dS model are partly due to the high
normalization in the model.
Observations of the local cluster abundance also place constraints on
the value of $\sigma_8$ which suggest low normalizations $\sigma_8\sim 
0.5$ to $0.6$ for the flat model with $\Omega_0=1$ (e.g., Eke, Cole \&
Frenk 1996; Kitayama \& Suto 1996, 1997; Viana \& Liddle 1996).
The E-dS model might be, therefore, considered as an extreme case.  

\section{EFFECT OF THE MAGNIFICATION BIAS ON QUASAR
         LUMINOSITY FUNCTIONS} 

Strictly speaking, almost all cosmological observations are under the
influence of the lensing magnification,
but there are very large differences in degree;
the majority of field distant sources (say, redshift $z_s>1$) might be
demagnified slightly (Wambsganss et al.\ 1998), whereas strongly
lensed sources such as multiply-imaged quasars and
giant luminous arcs, which are relatively very rare, 
are highly magnified (see e.g. SEF; and for a recent review, 
Mellier 1999).

In this section, we study the effect of the magnification bias 
on quasar luminosity functions, using the magnification
distributions obtained from the ray-tracing experiments described in 
\S\S2 and 3.
Although magnification bias is not a phenomenon restricted to quasars alone,
we focus our attention on these sources, as previous authors
have done (e.g., Schneider 1992), because lensing 
effects are most relevant to them, due to their inferred
compactnesses and large cosmological distances.

We use the conventional definition of the luminosity function 
$\Phi(L,z)$ of quasars, as being the
comoving number density of quasars at a redshift $z$ with a
luminosity $L$. We designate by $\Phi_{\rm int}(L,z)$ the
{\it intrinsic luminosity function} of quasars, which is the actual 
luminosity function, and by $\Phi_{\rm obs}(L,z)$ the {\it observed 
luminosity function}, which is the luminosity function inferred
from observations. Observers compute $\Phi_{\rm obs}(L,z)$ by counting
sources in redshift bins, and estimating their luminosities from their 
apparent brightnesses and the luminosity distance $D_L(z)$ corresponding to 
each redshift bin. In general, $\Phi_{\rm obs}(L,z)$ and $\Phi_{\rm int}(L,z)$
differ in presence of gravitational lensing. A source believed to have 
a luminosity $L$ might actually be a source with luminosity $L/\mu$
which is magnified\footnote{or demagnified, since $\mu$ can be 
either larger or smaller than unity.} by a factor $\mu$.
If we neglect any other uncertainty
in the determination of $\Phi_{\rm obs}$, then $\Phi_{\rm obs}(L,z)$ and 
$\Phi_{\rm int}(L,z)$ are related by
\begin{equation}
\label{obs-LF}
\Phi_{\rm obs}(L,z) = \int_0^{\infty} d\mu\, 
\Phi_{\rm int}\left({L\over\mu},z\right) p_s (\mu, z)
\end{equation} 

\noindent (SEF, chapter 12). In the absence of lensing, that is, in a
smooth Friedmann universe, $p_s(\mu,z)$ is equal to a Dirac $\delta$-function
$\delta_D(\mu-1)$, and equation (\ref{obs-LF}) reduces to 
$\Phi_{\rm obs}=\Phi_{\rm int}$.
We define the {\it bias parameter} $b_Q$ as
\begin{equation}
\label{def_bQ}
b_Q \equiv{\Phi_{\rm obs}(L,z) \over \Phi_{\rm int}(L,z)}\,.
\end{equation}

\noindent To compute the magnification bias, we need two ingredients:
the magnification probability $p_s(\mu,z)$ and the intrinsic luminosity
function $\Phi_{\rm int}(L,z)$. The former is provided by our ray-tracing 
experiments. For the latter, we consider two empirical models:
a single power-law model, and a double power-law model.

\subsection{Single Power-law Model}

Let us first consider an intrinsic luminosity function 
described by a single power-law model,
\begin{equation}
\label{single-LF}
\Phi_{\rm int}(L,z) = \Phi^{\ast}(z) \left( {L \over {L^{\ast}}}
\right)^{- \alpha}\,,
\end{equation}

\noindent
where the amplitude $\Phi^{\ast}(z)$ depends on redshift but not luminosity.
The power index $\alpha$ of the luminosity function is uncertain, but is
believed to be in the range $\alpha\sim1.2-1.6$ at the faint end of the
luminosity function, and $\alpha\sim3-4$ at the bright end
(Boyle et al.\ 1988; La Franca \& Cristiani 1997).
Hence, a single power-law model is not a particularly good 
approximation for $\Phi_{\rm int}$. However, this model
is interesting from a theoretical viewpoint.

We substitute equation (\ref{single-LF}) in equation ({\ref{obs-LF}), and get
\begin{equation}
\label{obs-LF-int}
\Phi_{\rm obs}(L,z) = \Phi_{\rm int}(L,z) \int_0^{\infty} d\mu\,\mu^\alpha
p_s (\mu, z)\,.
\end{equation} 

\noindent In this case, the bias parameter is
\begin{equation}
\label{def_bQ2}
b_Q= \int_0^{\infty} d\mu\,\mu^{\alpha} p_s (\mu, z)\,.
\end{equation}

\noindent Notice that $b_Q$ is a function of $z$, but not $L$. 
Hence, in the case of the single
power-law model, the logarithmic slope of the luminosity function is
unchanged by the magnification bias, but the amplitude is multiplied
by a redshift-dependent factor, and can either
increase or decrease, depending on 
the power index $\alpha$ and the probability distribution
$p_s(\mu,z)$. Notice also that in the particular
case of a power-law index 
$\alpha=1$, we get $\Phi_{\rm obs}=\Phi_{\rm int}$. The observed luminosity
functions in a smooth Friedmann universe and
in an inhomogeneous Friedmann universe are then equal, in spite of the
presence of lensing magnification.

\begin{figure}
\begin{center}
\begin{minipage}[t]{8cm}
\begin{center}
\epsfig{figure=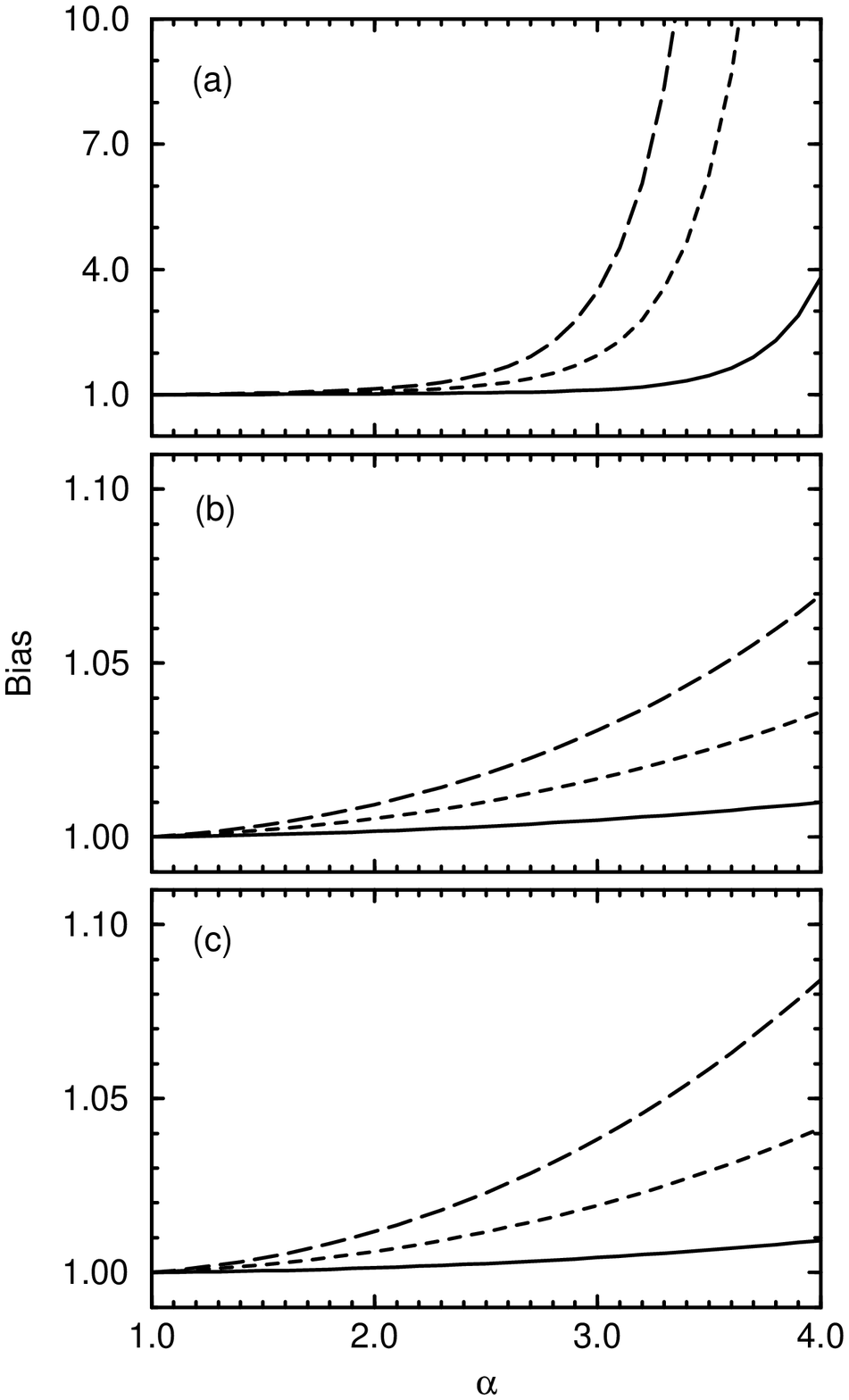,width=8cm}
\caption[]{
The bias parameter $b_Q$ as a function of a power index $\alpha$.
The intrinsic luminosity function is taken by the single power-law
form, equation (\ref{single-LF}). 
The solid lines are for the source redshift $z_s=1$, dashed lines are
for $z_s=2$, and long-dashed lines are for $z_s=3$.
(a) E-dS model; (b) O model; (c) $\Lambda$ model.
\label{fig7}}
\end{center}
\end{minipage}
\hspace{0.2cm}
\begin{minipage}[t]{8cm}
\begin{center}
\epsfig{figure=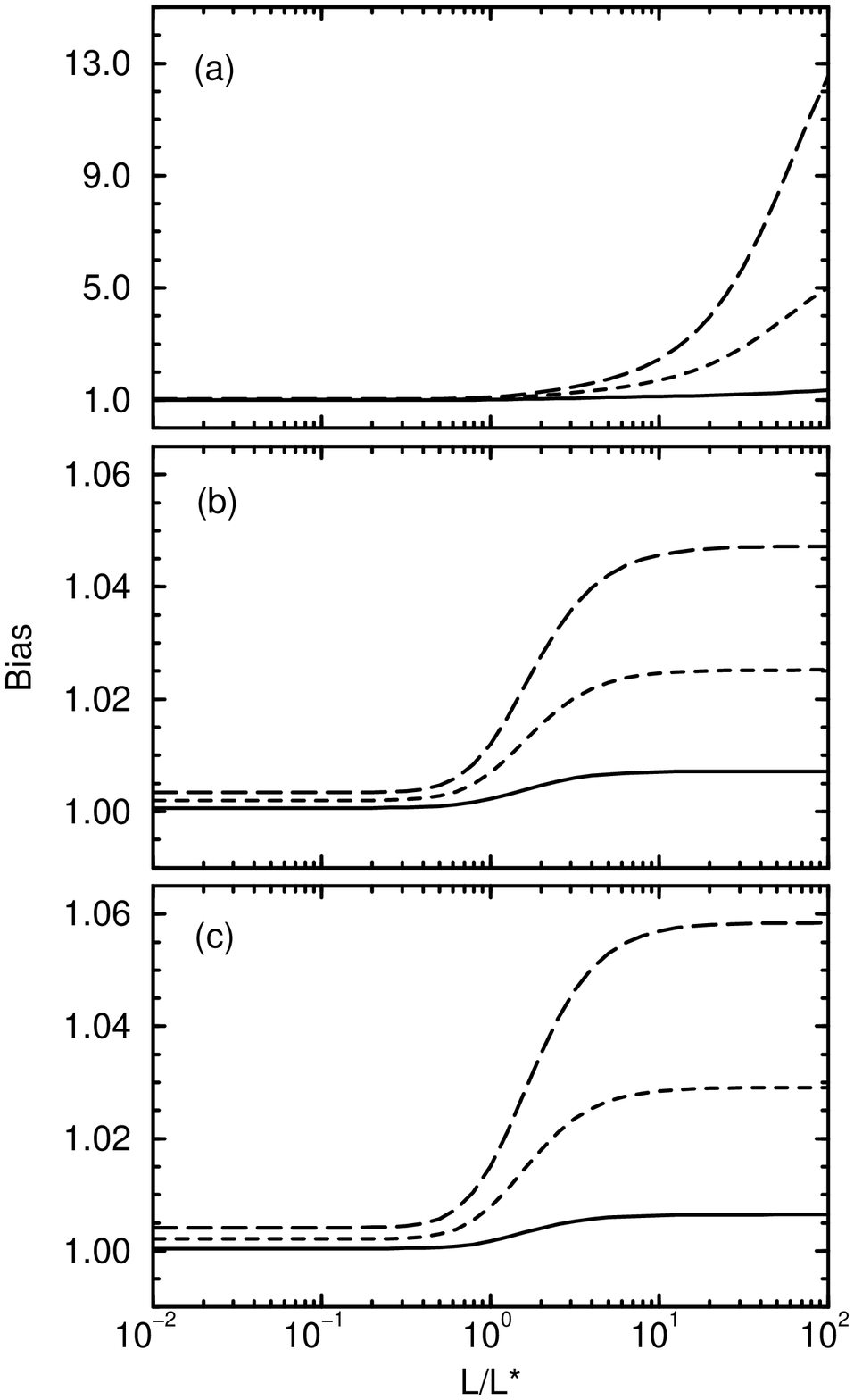,width=8cm}
\caption[]{
The bias parameter $b_Q$ as a function of a luminosity normalized by
its characteristic value, $L/L^{\ast}$. 
The intrinsic luminosity function is taken by the double power-law
form, equation (\ref{double-LF}), with $\alpha_1=1.5$, and $\alpha_2=3.5$.
The meanings of lines are the same as in Figure \ref{fig7}.
(a) E-dS model; (b) O model; (c) $\Lambda$ model.
\label{fig8}}
\end{center}
\end{minipage}
\end{center}
\end{figure}

We computed the bias parameter $b_Q$ as a function of the power-law index
$\alpha$, using the probability distribution functions derived from our
ray-tracing experiments. As in \S3.2, we limited the
integration range to $0.1\leq\mu\leq100$. We considered three particular
source redshifts, $z_s=1$, $z_s=2$, and $z_s=3$.
The results are plotted in Figure \ref{fig7}.
The bias strongly depends on the power index $\alpha$.
Since the actual quasar luminosity function becomes steeper as the luminosity
increases, the effect of
magnification bias should be more important at the bright end
of the luminosity function than at the faint one.
Figures \ref{fig7}b and \ref{fig7}c show that the biasing
effect is moderate for the O and $\Lambda$ models, and is at most
$10\%$ even for sources at $z_s=3$.
On the other hand, the biasing is considerable for the E-dS model.
Even for sources at moderate redshift, $z_s=1$, the bias parameter
$b_Q$ is 4 for $\alpha=4$.
For sources at $z_s=2$ and $3$, the bias parameter
strongly grows with $\alpha$ for the E-dS model. 
This is mainly due to the presence of
the power-law tail in the magnification distributions that we
saw in Figure \ref{fig5}a.
When the power-law tail appears in the magnification probability,
the bias parameter scales roughly
as $b_Q \propto \mu_{\max}^{\alpha-2}$, where $\mu_{\max}$ is the upper
value of the integration over $\mu$, in our case $\mu_{\max}=100$.
This implies that the magnification bias diverges in the limit of
large magnifications, corresponding to low luminosities, when
$\alpha>2$. This divergence is essentially an artifact of the single power-law
model used here. The actual luminosity function flattens 
at the faint end, and the power-law index drops below 2, thus
eliminating the divergence.
Furthermore, the finite size of the source size becomes an important effect
for very highly magnified sources (say $\mu >10^3$, Schneider 1987c, 1992).
Schneider \& Weiss (1988b) and Schneider (1992) have pointed out that
the magnification distribution is effectively cut off like
$p(\mu)\propto (\mu_c/\mu)^6$ for sources with a finite extend, where
$\mu_c$ is a function of the source extend and the mass of a lens.
In the case we considered (i.e. distant quasars lensed by large-scale 
structure), $\mu_c$ is larger than $10^3$ (see, e.g., SEF, chapter 12).
We do not need to take this effect into account here, 
because the upper limit of the integration, $\mu_{\max}$, is smaller 
than $\mu_c$.

\subsection{Double Power-law Model}

Since the power-law index of the luminosity functions appears to take
different values at the bright and faint ends, a single power law model, 
such as the one considered in \S4.1, does not provide a very good fit, for 
any value of the power-law index $\alpha$. A much better fit can be obtained
by considering a double power-law model of the form
\begin{equation}
\label{double-LF}
\Phi_{\rm int}(L,z) = {{\Phi^{\ast}(z)} \over 
{ \left( {L/ {L^{\ast}}} \right)^{\alpha_1} + 
\left( {L/ {L^{\ast}}} \right)^{\alpha_2}}},
\end{equation}

\noindent
where $\alpha_1$ and $\alpha_2$ are the logarithmic slope of the faint
and bright end sides of the luminosity function, respectively.
We choose $\alpha_1 =1.5$, and $\alpha_2=3.5$ which are typical values 
(Boyle et al.\ 1988; La Franca \&
Cristiani 1997).

In Figure \ref{fig8}, we plot the biasing parameter $b_Q$ 
in units of its characteristic value $L^{\ast}$. 
Figures \ref{fig8}b and \ref{fig8}c show that the 
magnification bias is moderate for O and $\Lambda$ models.
The bias enhances the luminosity functions by about $5\%$ at most at the bright
end, and has little effect (below $1\%$) at the faint end.
On the other hand, the effect is significant in the E-dS model,
especially for sources at high redshift, $z_s >2$.
At the present time, high-$z$ quasars seems to be homogeneously
sampled to 2 or 3 magnitude brighter than their characteristic value
$M^{\ast}$ (which corresponds to a luminosity $L\sim10L^{\ast}$,
e.g., La Franca \& Cristiani 1997).
Figure \ref{fig8}a shows that, for the E-dS model, the bright
end of luminosity function (at $L \sim 10 L^{\ast}$) is enhanced by
a factor larger than 2 for such high-$z$ quasars.
In this case, the observed luminosity function of
bright quasars is strongly biased by the
lensing magnification effect.

\begin{figure}[t] 
\begin{center}
\begin{minipage}{9cm} 
\epsfig{figure=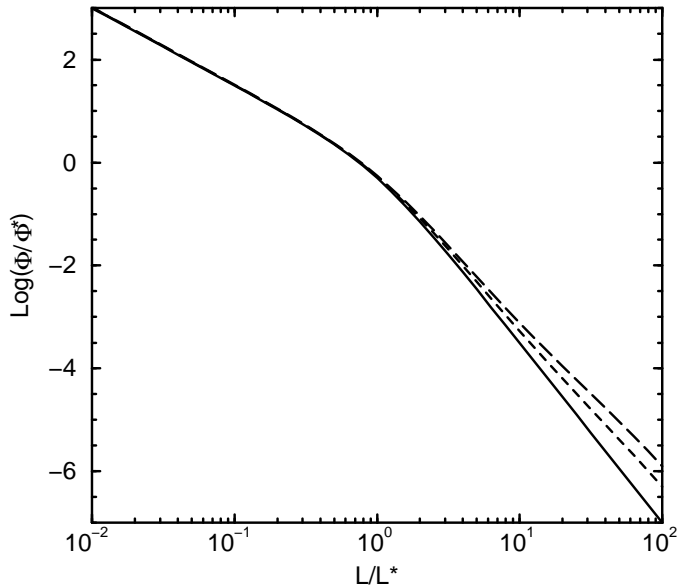,width=9cm}
\end{minipage}
\end{center}
\caption[]{
The observed luminosity function at the redshifts $z=2$ (the dashed
line), and $z=3$ (the long-dashed line) are plotted as a function of
the luminosity normalized by its characteristic value, $L/L^{\ast}$,
together with the intrinsic luminosity function (the solid line).
The intrinsic luminosity function is taken by the double power-law
form, equation (\ref{double-LF}) with the power indices $\alpha_1=1.5$,
and $\alpha_2=3.5$.
The background cosmology is taken by E-dS model. \label{fig9}}
\end{figure}

As an illustrative example, we plot in Figure \ref{fig9},
for the E-dS model, the observed luminosity function
as a function of the luminosity in units of its characteristic value
$L^{\ast}$, together with the intrinsic luminosity function,
for sources at redshifts $z_s=2$ (dashed line) and at $z_s=3$ 
(long-dashed line). The lensing bias flatten the slope of
the luminosity functions at the bright end of the luminosity
function, above the characteristic luminosity $L^{\ast}$.
If we fit the observed luminosity functions using the same double power-law
model (eq. [\ref{double-LF})]) as for the intrinsic luminosity function,
we find that the values of $L_{\ast}$ and $\alpha_1$ are the same,
but the ``observed'' power-law index $\alpha_2$ takes the values 
$\alpha_{2,\rm obs}=3.0$, and $\alpha_{2,\rm obs}=2.8$ for $z_s=2$ and 3, 
respectively, compared to $\alpha_2=3.5$ for the intrinsic luminosity function.
Since the observed power-law index depends on the source redshift, it
is essential to take into account the lensing magnification bias when
studying the evolution of the {\it intrinsic} quasar luminosity functions,
in order to untangled the effects of magnification bias and genuine
luminosity and/or number evolution.

\section{SUMMARY AND CONCLUSION}

We have studied the statistical properties of gravitational lensing 
by large-scale structure, for three different {\it COBE}-normalized
cosmological models.
We used a P${}^3$M $N$-body code to simulate the formation and evolution of 
large-scale structure, and then used the multiple lens-plane 
algorithm to follow light rays propagating through the inhomogeneous
matter distribution in the universe. 
For each model, we followed the propagation of $1.1\times10^7$
light rays, from the observer up to a source redshift of 3, and
computed the evolution of the Jacobian matrix along each ray. 
Having such a large number of light rays
enabled us to compute, as a function of source redshift, the
distributions and rms values of the lensing convergences, shears, and
magnifications, to high accuracy. 
We compared these results with various analytical estimates
based on the power spectrum method.
Finally, by combining the magnification probability obtained from the
experiments with empirical models for the intrinsic luminosity
function, of quasars, we computed the magnification bias.
Our main results can be summarized as follows:

\begin{enumerate}
\item
The rms values of the lensing convergences and shears, which are
expected to be equal according to analytical arguments,
were found to be nearly equal,
the difference being attributed to the finite resolution of the 
numerical methods used for the simulations. These values were
consistent with analytical predictions based on
the power spectrum approach with nonlinear evolution of
the power spectra.
\item
The dispersion of the lensing magnifications strongly depends on
the amplitude of the density fluctuations of the matter and thus
depends on the density parameter $\Omega_0$.
The dispersion becomes larger than 0.1 at $z_s=1$ for the E-dS model, and
at $z_s=3$ for the O and $\Lambda$ models. 
The magnification distributions are considerably skewed toward high
magnification. In particular, a power-law tail appears at large
magnifications in the E-dS model, for sources at redshifts $z_s >1$.
\item
We have compared the statistics of the lensing magnification with the
nonlinear predictions of the power spectrum approach.
We found that the power spectrum approach with the nonlinear power
spectrum correctly predicts the value of the dispersion of the magnification
in the absence of the power-law tail. 
However, once the power-law tail appears, the predicted values become
inaccurate, because the dispersions grow
more rapidly with the source redshift than the nonlinear predictions.
These predictions, therefore, should be regarded as a lower limit.
\item
We studied the magnification bias on quasar luminosity functions.
We found that the lensing magnification bias strongly depends on the
slope of the luminosity function and on the density parameter $\Omega_0$. 
If the mean matter density is as high 
as in the E-dS model, the biasing effects can be significant, especially at
the bright end of the luminosity function where its slope is very
steep. 
Moreover, since quasar luminosity functions becomes steeper with the
luminosity, the lensing bias will flatten the effective slope of an
observed luminosity functions at the bright side.
\end{enumerate}

The power-law tail did not appear in the O and $\Lambda$ models
because the large-scale structure in these models is not as
evolved as in the E-dS model.
However, the existence of strongly
magnified images, such as multiply imaged quasars or galaxies and
giant luminous arcs, suggests that whatever the background cosmological
model is, there is a power-tail in the actual magnification distribution.
Therefore, in order to study the effect of lensing magnification
bias on the observed luminosity function, one must carefully take
into account the influence of this power-law tail.
It is also important to emphasize that, as we have stated in
\S3.2, the large lensing effects found in the E-dS
model are partly due to the high normalization used for that model.
Therefore, the E-dS model should be considered as an extreme case.

We have investigated the magnification bias resulting from the 
gravitational lensing in universes with realistic matter distributions.
In order to study quantitatively the biasing effects on the
observed quasar luminosity functions, one has to take into account
additional selection effects such like the finite extend of a source, 
which have not been considered in this paper.
Such effects will be studied in future works.

\acknowledgments
We would like to thank P. Premadi for useful discussions.
This research was supported in part by the Grants-in-Aid by the Ministry 
of Education, Science, Sports and Culture of Japan (09640332).
HM was supported by NASA Grants NAG5-2785 and NAG5-7363, and
a fellowship from the Texas Institute for Computational and Applied 
Mathematics. The P$^3$M simulations were performed at the High 
Performance Computing Facility, University of Texas.

\appendix

\section{CALCULATION OF THE VARIANCES}

In this Appendix, we present the analytic formulae for calculating the
variances of the lensing convergences, shears, and magnifications.
Those formulae have been derived by Kaiser (1992, 1998), Villumsen
(1996), Bernardeau et al. (1997), Frieman (1997),
Jain \& Seljak (1997), and Nakamura (1997). We refer the reader to 
these references for details, and only present the final expressions.

In the weak gravitational field limit, 
the variance of the lensing convergence and shear can be estimated using
both the Born approximation (Schneider et al. 1998) 
and Limber's equation in Fourier space (Kaiser 1992; 1998). 
These variance, which turn out to be equal,
are related to the density power spectrum, by
\begin{equation}
\label{ap:sigma_k}
\sigma_{\kappa}^2 [a(\chi)]=\sigma_{\gamma}^2 [a(\chi)]
={{9 \Omega_0^2} \over {8 \pi}} \left( {{H_0 \over c}} \right)^4 
\int_0^{\chi} dv \left[{D(0,v) D(v,\chi) \over a(v)D(0,\chi)} \right]^2
I(v)\,,
\end{equation}

\noindent
where $\chi$ is the {\it comoving distance}, 
$D(\chi_1,\chi_2)$ is the standard angular diameter distance between $\chi_1$ 
and $\chi_2$, and $a$ is the Robertson-Walker
scale factor normalized to be unity at the
present. The relationship between the comoving distance and the
redshift $z$ can be derived from the Friedmann equation (Jain \&
Seljak 1997): 
\begin{equation}
\label{lambda-z}
\chi (z) = {c \over {H_0}} \int_{1 /(1+z)}^1 da \left[\lambda_0 a^4 
+(1-\Omega_0-\lambda_0) a^2 + \Omega_0 a \right]^{-{1/2}}.
\end{equation}

\noindent
The function $I(v)$ appearing in equation (A1) in an integral over the
power spectrum. It is the particular form of that integral that distinguishes
the various approximations. In linear theory, this function is given by
\begin{equation}
I(v)=\int_0^\infty dk\,kP_{\rm L}[k,a(v)]\,,
\end{equation}

\noindent where $P_{\rm L}$ is the linear power spectrum, which is a function
of the wavenumber $k$ and the scale factor $a$, and the dependence of the 
comoving distance $v$ enters only through the scale factor. 
In the nonlinear approximation, we use instead
\begin{equation}
I(v)=\int_0^\infty dk\,kP_{\rm NL}[k,a(v)]\,,
\end{equation}

\noindent where $P_{\rm NL}$ is the nonlinear power spectrum, for which
Peacock \& Dodds (1996) provide a fitting formula. Finally, for the
top-hat filtered density field approximation, we use
\begin{equation}
I(v,R_s)=\int_0^\infty dk\,kP_{\rm NL}[k,a(v)]\hat W^2_{\rm TH}(kR_s)\,,
\end{equation}

\noindent 
where $\hat{W}_{\rm TH}(x)\equiv3 (\sin x
-x \cos x)/x^3$ is the Fourier transform of the top-hat window
function with smoothing scale $R_s$.
The expression of the variance of the lensing magnifications have been
derived by Frieman (1997) and Nakamura (1997), who found that
$\sigma_{\mu}^2= 4\sigma_{\kappa}^2$.


\end{document}